
\let\includefigures=\iftrue
%
\let\includefigures=\iffalse
%
\let\useblackboard=\iftrue
%
%
%
\input harvmac.tex
\message{If you do not have epsf.tex (to include figures),}
\message{change the option at the top of the tex file.}
\input epsf
\epsfverbosetrue
\def\fig#1#2{\topinsert\epsffile{#1}\noindent{#2}\endinsert}
\def\fig#1#2{}
%
\def\Title#1#2{\rightline{#1}
\ifx\answ\bigans\nopagenumbers\pageno0\vskip1in%
\baselineskip 15pt plus 1pt minus 1pt
\else
\def\listrefs{\footatend\vskip 1in\immediate\closeout\rfile\writestoppt
\baselineskip=14pt\centerline{{\bf References}}\bigskip{\frenchspacing%
\parindent=20pt\escapechar=` \input
refs.tmp\vfill\eject}\nonfrenchspacing}
\pageno1\vskip.8in\fi \centerline{\titlefont #2}\vskip .5in}

\ifx\answ\bigans\def\tcbreak#1{}\else\def\tcbreak#1{\cr&{#1}}\fi
\useblackboard
\message{If you do not have msbm (blackboard bold) fonts,}
\message{change the option at the top of the tex file.}
\font\blackboard=msbm10 scaled \magstep1
\font\blackboards=msbm7
\font\blackboardss=msbm5
\newfam\black
\textfont\black=\blackboard
\scriptfont\black=\blackboards
\scriptscriptfont\black=\blackboardss

\else

\fi
%
\def\yboxit#1#2{\vbox{\hrule height #1 \hbox{\vrule width #1
\vbox{#2}\vrule width #1 }\hrule height #1 }}
\def\fillbox#1{\hbox to #1{\vbox to #1{\vfil}\hfil}}
\def\ybox{{\lower 1.3pt \yboxit{0.4pt}{\fillbox{8pt}}\hskip-0.2pt}}
\def\comments#1{}

\def\p{\partial}

\Title{\vbox{\baselineskip12pt
\hfill{\vbox{
\hbox{BROWN-HET-1027\hfil}
\hbox{hep-th/9512042}}}}}
{\vbox{\centerline{Dirichlet Boundary State in Linear Dilaton Background}
\vskip20pt
\centerline{}}}
\centerline{Miao Li}
\smallskip
\centerline{Department of Physics}
\centerline{Brown University}
\centerline{Providence, RI 02912}
\centerline{\tt li@het.brown.edu}
\bigskip
\noindent

Dirichlet-branes have emerged as important objects in studying
nonperturbative string theory. It is important to generalize these objects
to more general backgrounds other than the usual flat background. The simplest
case is the linear dilaton condensate. The usual Dirichlet boundary
condition violates conformal invariance in such a background. We show
that by switching on a certain boundary interaction, conformal invariance
is restored. An immediate application of this result is to
two dimensional string theory.

\Date{December 1995}
\nref\joe{J. Polchinski, ``Dirichlet-Branes and Ramond-Ramond Charges,''
hep-th/9510017.}
\nref\witten{E. Witten, ``Bound States of Strings and p-Branes,''
hep-th/9510135.}
\nref\miao{M. Li, ``Boundary States of D-Branes and Dy-Strings,''
hep-th/9510161, to appear in Nucl. Phys. B.}
\nref\ck{C. G. Callan and I. R. Klebanov, ``D-Brane Boundary State
Dynamics,'' hep-th/9511173.}
\nref\sen{A. Sen, ``A Note on Marginally Stable Bound State in Type II
String Theory,'' hep-th/9510225; ``U-duality and Intersecting
D-branes,'' hep-th/9511026.}
\nref\kt{I. R. Klebanov and L. Thorlacius, ``The Size of p-Branes,''
hep-th/9510200; C.~Bachas, ``D-Brane Dynamics,'' hep-th/9511043.}
\nref\joep{J. Polchinski, Phys. Rev. D50 (1994) 6041.}
\nref\lm{M. Li and S. Mathur, work in progress.}
\nref\clny{C. G. Callan, C. Lovelace, C. R. Nappi and S. A. Yost, Nucl.
Phys. B308 (1988) 221.}
\nref\cklm{C. G. Callan and I. R. Klebanov, Phys. Rev. Lett. 72 (1994)
1986; C.~G.~Callan, I.~R.~Klebanov, A.~W.~W.~Ludwig and
J.~Maldacena, Nucl. Phys. B422 (1994) 417.}
\nref\steve{S. Shenker, `` Another Length Scale in String Theory?''
hep-th/9509132.}
\nref\gl{M. Goulian and M. Li, Phys. Rev. Lett. 66 (1991) 2051.}

Polchinski's D-branes \joe\ represent an exact treatment of various p-brane
solutions. Such objects become increasingly important in understanding
nonperturbative aspects of string theory, in particular duality. The
construction was immediately generalized to include bound states \witten\
\sen, and boundary states associated to them are studied in \miao\ and
\ck. The stringy properties of D-branes are explored in \kt. All these
are perfectly in harmony with duality conjectures.

Properties of D-branes in a curved background have not been examined
in much detail. The simplest nontrivial background is the linear
dilaton condensate, which occurs in many interesting situations
such as two dimensional string theory, and two dimensional black hole
and related models. As being already pointed out in \joep, the simple
Dirichlet boundary condition does not respect conformal invariance
in presence of a background charge.
A mechanism proposed there is to use degenerate loops to compensate
such violation. We shall adopt another strategy in this note to
restore conformal invariance: Adding a certain boundary term. This
may be connected to suggestion of \joep, although we have not
tried to explore this possibility. The necessity of such a boundary
term indicates that in order to include D-branes in a linear dilaton
background, certain open string background must be switched on.
Perturbative closed strings do not sense the existence of such background,
since diagrams involved are Riemann surfaces without boundaries.
Some applications of our result to two dimensional string are being worked
out in \lm.

For simplicity, we consider only one free scalar on the world-sheet
denoted by $\phi$. The holomorphic component of the stress tensor reads
\eqn\stress{T=-{1\over 2}(\p\phi)^2+Q\p^2\phi,}
and a similar anti-holomorphic counterpart. The central charge of
this free scalar is $c=1+12Q^2$, and $Q=\sqrt{2}$ in two dimensional
string theory. Consider a unit disk, the conformal invariance condition
on the boundary is $Tdz^2=\bar{T}d\bar{z}^2$. In other words, there
is no net energy-momentum flow out of the boundary in world-sheet point
of view. It is convenient to work with mode expansions
\eqn\mode{\eqalign{\phi &=\varphi_0-ip(\hbox{ln}z+\hbox{ln}\bar{z})
-i\sum_{n\ne 0}{1\over n}\left(\alpha_{-n}z^n+\tilde{\alpha}_{-n}
\bar{z}^n\right),\cr
L_n &=[p+iQ(n+1)]\alpha_n+{1\over 2}\sum_{m\ne 0}\alpha_{m+n}\alpha_{-m},}}
a similar formula for $\tilde{L}_n$. The commutators are $[\alpha_m,
\alpha_n]=m\delta_{m+n,0}$, and similarly for the right-moving modes.
Let $K_n=L_n-\tilde{L}_{-n}$. The boundary condition is entirely
encoded in the boundary state $|B\rangle$, and the conformal invariance
condition is $K_n|B\rangle=0$.

The usual Neumann boundary condition is given by $\partial_r\phi=0$
on the boundary of the unit disk. In terms of the boundary state,
it states that
$$p|B\rangle_N=(\alpha_n+\tilde{\alpha}_{-n})|B\rangle_N
=0.$$
Due to the existence of the background charge $Q$, one has to modify
the boundary condition a bit: $p=-iQ$. So there must be a net momentum
flow out of the boundary (in view of spacetime $\phi$). One way to
see this is to consider the commutators
\eqn\ocomm{[K_m, \alpha_n+\tilde{\alpha}_{-n}]=2m(p+iQ)\delta_{m+n,0}
-m\left(\alpha_{m+n}+\tilde{\alpha}_{-m-n}\right).}
So when $p=-iQ$, the center term disappears, and it is possible to impose
both the conformal invariance condition and Neumann boundary
condition. Dirichlet boundary condition is $\p_\theta\phi=0$
on the boundary.
As Polchinski already observed \joep, this simple condition is
no longer conformally invariant if $Q\ne 0$. Or under a general
conformal transformation, $\delta\phi$ acquires a term proportional
to the Weyl factor generally nonvanishing on the boundary. Another way to
see this is to derive similar commutators as \ocomm. In terms of modes,
Dirichlet boundary condition is
$$(\alpha_n-\tilde{\alpha}_{-n})|B\rangle_D=0.$$
This is not compatible with the conformal invariance condition since
\eqn\scomm{[K_m, \alpha_n-\tilde{\alpha}_{-n}]=2im^2Q\delta_{m+n,0}
-m\left(\alpha_{m+n}-\tilde{\alpha}_{-m-n}\right).}
The center term, independent of $p$, is always nonvanishing.

To restore conformal invariance, we have to modify the boundary
condition. To be as close to the ordinary Dirichlet condition as
possible, one requires that a net momentum transfer is possible
if one scatters string states against the object described by
the boundary state. So $|B,p\rangle$ is an eigen-state of $p$ with
arbitrary number $p$. To solve equations $K_n|B,p\rangle=0$,
it is convenient to adopt the coherent state technique introduced
in \clny. Introduce the following coherent states
\eqn\cohe{(\alpha_n-\tilde{\alpha}_{-n}-x_n)|x,p\rangle=0,}
where $n$ can be either positive or negative. The Hermiticity
condition $x_{-n}=\bar{x}_n$ must be met.
This set of states forms a complete orthogonal basis.
The solution to \cohe\ is
\eqn\coher{|x,p\rangle=\exp\left(\sum_{n=1}^\infty{1\over n}[-{1\over 2}
x_nx_{-n}+\alpha_{-n}\tilde{\alpha}_{-n}+x_n\alpha_{-n}-
x_{-n}\tilde{\alpha}_{-n}]\right)|p\rangle.}
It can be checked that these states normalize to delta function.

We postulate that the desired modified Dirichlet boundary state
is of the form
\eqn\post{|B,p\rangle=\int [dx]|x,p\rangle\Phi(x).}
To solve $K_n|B,p\rangle=0$, one first computes
$$\eqalign{K_n|x,p\rangle &=(2iQn\tilde{\alpha}_{-n}+\sum_{m>0}
[x_{m+n}\alpha_{-m}+\bar{x}_{m-n}\tilde{\alpha}_{-m}]+(p+iQ(n+1))x_n
\cr
&+{1\over 2}\sum_{0<m<n}x_{n-m}x_m) |x,p\rangle}$$
for $n>0$. A similar formula can be derived for $n<0$. Observing
the form \coher, one replaces $\alpha_{-m}$ in the above formula
by $m\p_m+{1\over 2}x_{-m}$
and $\tilde{\alpha}_{-m}$ by $-m\p_{-m}-{1\over 2}x_m$. Thus $K_n$
is replaced by a first order differential operator when acts on
$|x,p\rangle$. Substituting this relation into \post\ and integrating
by parts, the conformal invariance condition for $\Phi$ is obtained
\eqn\diffe{\left(2iQn^2\p_n +(p+iQ)x_{-n}-  \sum_{m=-\infty}^\infty m
x_{m-n}\p_m\right)\Phi(x)=0,}
incidentally this is valid for both $n>0$ and $n<0$. Assume $\Phi=e^K$,
the differential equations for $K$ follow from \diffe\
\eqn\diffk{2iQn^2\p_n K+(p+iQ)x_{-n}= \sum_{m=-\infty}^\infty m
x_{m-n}\p_m K.}

We now solve differential equations \diffk. Observe that these equations
can be solved recursively. Let $K=\sum_N K_N$, where $K_N$ contains
N-th powers in the $x$'s. The constant term is not much of interest
at present. There can be no linear term. The first term is $K_2$
satisfying
\eqn\sec{2iQn^2\p_n K_2+(p+iQ)x_{-n}=0,}
with solution
\eqn\ssec{K_2=-(p+iQ)({i\over 2Q})\sum_{m,n}{1\over 2mn}x_mx_n\delta_{m+n,0}.}
The recursive relation is then
\eqn\recur{2iQn^2\p_nK_{N+1}=\sum m x_{m-n}\p_mK_N.}
The ansatz
$$K_N=a_N\sum_{m_i}{1\over m_1\dots m_N}x_{m_1}\dots x_{m_N}
\delta_{m_1+\dots+m_N,0}$$
leads to
$$a_{N+1}={i\over 2Q}{a_N\over N+1}=-(p+iQ)({i\over 2Q})^N{1\over (N+1)!}.$$
Remarkably, the sum $\sum_N K_N$ is given by a very simple
form, up to a constant term
\eqn\solut{
\eqalign{K&=2iQ(p+iQ)\oint {d\theta\over 2\pi}e^{-{1\over 2Q}X(\theta)},\cr
X(\theta)&=-i\sum_m {1\over m}x_me^{im\theta}.}}
$X(\theta)$ is real-valued. When only a pair $x_n$ and $x_{-n}$ are
nonvanishing, $X(\theta)$ is a sine function in $\theta$ and the phase
of $x_n$.

Finally, the boundary state is given by
\eqn\fin{|B,p\rangle=\int [dx]|x,p\rangle\exp\left(2iQ(p+iQ)\oint
{d\theta\over 2\pi}e^{-{1\over 2Q}X(\theta)}\right).}
The following remarks on \fin\ are in order. The solution to the
infinite set of differential equations \diffk\ is by no means unique.
However, we trust that the solution given by \fin\ is the appropriate
generalization of the usual Dirichlet boundary state to the background
of linear dilaton condensate, because not only the solution looks
very elegant, but also it appears to return to the usual Dirichlet
boundary state in the limit $Q\rightarrow 0$. In this limit, whenever
$X(\theta)\ne 0$, the exponent $\int d\theta \exp(-X/(2Q)$ is large,
so the integral in \fin\ tends to center at $X=0$ which is the usual
Dirichlet state. As a consistency check, take $p=-iQ$, then $\Phi=1$.
Integrating over $x$ we obtain the Neumann boundary state discussed
before. For
a real $Q$, this is unphysical if we are interested in real momentum
transfer.

As we have expected, the form of \fin\ tells us that a boundary operator
which is to replace the ``wave function'' $\Phi(x)$ is needed in
order to restore conformal invariance. What is a little surprising is
that the coefficient $2iQ(p+iQ)$ is fixed for a given $Q$ and $p$.
If one attempts to replace $x_m$ in $X(\theta)$ by $\alpha_m-
\tilde{\alpha}_{-m}$, one obtains operator $\phi$ without the zero
mode part. Again, integrating over the $x$'s results in the Neumann
boundary state, except that the zero mode part is that of Dirichlet.
We conclude that the generalized Dirichlet boundary state in a linear
dilaton background is obtained by applying a boundary operator
$$\exp\left(2iQ(p+iQ)\oint
{d\theta\over 2\pi}e^{{1\over 2Q}\phi_{oc}}\right)$$
to the Neumann boundary state carrying momentum $p$.
We note in passing that a similar interaction
boundary term is studied in \cklm, where no background charge is
introduced. The usual Dirichlet boundary state is achieved by letting the
coupling constant of the boundary interaction go to infinity. While
such a limit is achieved by taking $Q\rightarrow 0$ in our case.
One more interesting aspect
deserves mentioning. For a real $Q$, the Liouville field $\phi$ can be
formally viewed as having an imaginary radius $R=2iQ$. It is just this
pure imaginary radius appearing in $\Phi(x)$. Our derivation presented
in this note is brute force in nature. The result looks quite elegant, so
it appears that there is a direct derivation based on the usual
conformal technique.

That the oscillator part of the boundary state is Neumann, despite the
appearance of a boundary interaction which helps to suppress spread
of oscillators, indicates that the object is not point-like in the
transverse direction. This should have some interesting physical
implication. Linear dilaton is a generic feature in conical phase
transitions, and D-branes in such a background pose many interesting
questions, for example, a new length scale \steve.
Whether our boundary state will shed light on the question of a new
scale in string theory remains to see.

Some applications of our main result \fin\ will appear in \lm, here we
only make a few preliminary comments. The boundary state with
location at $\phi=\phi_0$ is obtained by Fourier transform
\eqn\fourier{|B,\phi_0\rangle=\int dpe^{-ip\phi_0}|B,p\rangle.}
Since the boundary wave function $\Phi$ involves $p$, the integration
over $p$ yields a delta function containing $\int d\theta \exp(-X/(2Q))$,
showing that the object is not point-like in $\phi$. The disk
amplitude, important in evaluating nonperturbative effects, is
given by
\eqn\disk{\langle B,\phi_0|0\rangle,}
or
\eqn\diskt{\langle B,\phi_0|\exp(-\mu\int d^2ze^{-\sqrt{2}\phi})|0\rangle,}
when a tachyon condensate is present. The second object is not
very easy to compute, and eventually one has to invoke the
analytic extension method of \gl. Generalization to some supersymmetric
situation is also under consideration, and there seems to be no intrinsic
difficulty in doing this.

\noindent {\bf Acknowledgments}

We would like to thank S. Mathur for collaboration and many useful
discussions, and A. Jevicki for helpful conversations. This work was
supported by DOE grant DE-FG02-91ER40688-Task A.

\listrefs

\end